\documentclass[apj]{emulateapj}
\usepackage{graphicx}
\usepackage{natbib}
\usepackage{upgreek}
\usepackage{color}

\newcommand{\kep}{$K_{ep}$}
\newcommand{\nh}{\bar{n}_{\rm H}}
\newcommand{\hii}{H$\,\scriptstyle{\rm II}$~}
\newcommand{\hi}{H$\,\scriptstyle{\rm I}$~}

\newcommand{\ddeg}{\hbox{$.\!\!^\circ$}}  
\newcommand{\masssun}{\mathrm{M_\odot}}
\newcommand{\mass}{\mathrm{M}}
\slugcomment{To be resubmitted to ApJ: v18.0,  \today}

\shortauthors{\emph{Fermi} LAT collaboration}
\title{\emph{Fermi} LAT Discovery of Extended Gamma-Ray Emissions in the Vicinity of the HB~3 Supernova Remnant}
\shorttitle{\emph{Fermi} LAT Discovery of Extended Gamma-Ray Emissions in the Vicinity of the HB~3 SNR}

\begin{document}


\author{
H.~Katagiri\altaffilmark{1,2},
K.~Yoshida\altaffilmark{1,2},
J.~Ballet\altaffilmark{3},
M.-H.~Grondin\altaffilmark{4},
Y.~Hanabata\altaffilmark{5},
J.W.~Hewitt\altaffilmark{6,7}, 
H.~Kubo\altaffilmark{8},
M.~Lemoine-Goumard\altaffilmark{4}
}
\altaffiltext{1}{College of Science, Ibaraki University, 2-1-1, Bunkyo, Mito 310-8512, Japan}
\altaffiltext{2}{Corresponding author: H.~Katagiri, hideaki.katagiri.sci@vc.ibaraki.ac.jp; K.~Yoshida, 13nm169s@gmail.com}
\altaffiltext{3}{Laboratoire AIM, CEA-IRFU/CNRS/Universit\'e Paris Diderot, 
Service d'Astrophysique, CEA Saclay, 91191 Gif sur Yvette, France}
\altaffiltext{4}{Centre d'\'Etudes Nucl\'eaires de Bordeaux Gradignan, 
IN2P3/CNRS, Universit\'e Bordeaux 1, BP120, F-33175 Gradignan Cedex, France}
\altaffiltext{5}{Institute for Cosmic-Ray Research, University of Tokyo, 
5-1-5 Kashiwanoha, Kashiwa, Chiba, 277-8582, Japan}
\altaffiltext{6}{Department of Physics and Center for Space Sciences and Technology, University of Maryland Baltimore County, Baltimore, MD 21250, USA}
\altaffiltext{7}{Center for Research and Exploration in Space Science and Technology (CRESST) and NASA Goddard Space Flight Center, Greenbelt, MD 20771, USA}
\altaffiltext{8}{Department of Physics, Graduate School of Science, Kyoto 
University, Kyoto, Japan}


\begin{abstract}
We report the discovery of extended gamma-ray emission measured by the Large Area Telescope~(LAT) onboard the \textit{Fermi Gamma-ray Space Telescope} in the region of the supernova remnant~(SNR) HB~3~(G132.7$+$1.3) and the W3 \hii complex adjacent to the southeast of the remnant.
W3 is spatially associated with bright $^{12}$CO~(J=1-0) emission. 
The gamma-ray emission is spatially correlated with this gas and the SNR. 
We discuss the possibility that gamma rays originate in interactions between particles accelerated in the SNR and interstellar gas or radiation fields.
The decay of neutral pions produced in nucleon-nucleon interactions between accelerated hadrons and interstellar gas provides a reasonable explanation 
for the gamma-ray emission. 
The emission from W3 is consistent with irradiation of the CO clouds by the cosmic rays accelerated in HB~3.

\end{abstract}


\section{Introduction}
Diffusive acceleration by supernova shock waves can  
accelerate particles to very high energies~\citep[e.g.,][]{blandford87}.
However, the processes of acceleration, release from the shock region, and diffusion in the interstellar medium of such particles has not been well understood so far.
Gamma-ray observations in the GeV regime are a powerful probe of these mechanisms.
The Large Area Telescope~(LAT) on board the \textit{Fermi Gamma-ray Space Telescope} has recently detected GeV gamma rays from SNRs~\citep[e.g., ][and references therein]{SNRref}.

 HB~3~(G132.7$+$1.3) is a well-known middle-aged SNR.
The 1.$\!\!^\circ$3 diameter~\citep{newRadioShape} makes it a good candidate
for detailed morphological studies in high-energy gamma rays since it is larger than the LAT angular resolution above 
{\bf 1~GeV (smaller than 1$^\circ$)}.
The W3 \hii complex is adjacent and to the southeast of the remnant.
Bright $^{12}$CO(J=1-0) line emission near $-$43~km~s$^{-1}$ around W3 is partly surrounded by a region of enhanced radio continuum emission from HB~3, suggesting that HB~3 is interacting with this gas~\citep{COint}.
The distance to the SNR is therefore considered to be that of W3, estimated to be 2.0--2.4~kpc~\citep[see e.g.][]{COint, dist4}. 
In this paper, we adopt 2.2~kpc.
The age was estimated to be $\sim$~$3 \times 10^4$~yr based on X-ray data and applying different evolutionary models to the SNR~\citep{age}. 
{\bf The X-ray spectrum was well described by a thermal plasma model~\citep{age}.}

In the COS B era, W3 was associated with the gamma-ray source CG135$+$1~\citep{COSB} and during the EGRET era it was associated with 3EG~J0229$+$6151~\citep{3eg}.  However, detailed spatial association and extension measurements could not be made due to the limited angular resolution of COS B and EGRET.
Five LAT sources positionally associated with HB~3 and W3 are listed in the 2FGL catalog~\citep{2yrCatalog}.
In this paper, we report a detailed analysis of \emph{Fermi} LAT observations around HB~3.
First, we briefly describe the observations and data selection in
Section~\ref{sec:obs}.
The analysis procedure and the results are presented in Section~\ref{sec:ana}, 
with the study of the morphology and spectrum of the emission associated with HB~3 and W3. 
Results are then discussed in Section~\ref{sec:discuss} 
and our conclusions are presented in Section~\ref{sec:conclusion}.

\section{OBSERVATIONS AND DATA SELECTION}
\label{sec:obs}
The LAT is the main instrument on \emph{Fermi}, detecting gamma rays from
$\sim$~20~MeV to $>$~300~GeV.
It is an electron-positron pair production telescope, built with tungsten foils and silicon microstrip detectors to measure the arrival directions of incoming gamma rays, and a hodoscopic cesium iodide calorimeter to determine the gamma-ray energies.
They are surrounded by 89 segmented  plastic scintillators that serve as an anticoincidence detector to reject charged particle events.
The on-orbit calibration, event classification and instrument performance are described in \cite{onorbit_cal} and \cite{IRFs}.
Compared to earlier high-energy gamma-ray telescopes, the LAT has a larger field of view~($\sim$~2.4~sr), a larger effective area~($\sim$~8000~cm$^2$ for $>$~1~GeV on-axis) and improved point-spread function~(PSF; the 68\% containment angle above 1~GeV is smaller than 1$^\circ$). 

Routine science operations with the LAT began on August 4, 2008.
We have analyzed events in the region of the HB~3 SNR collected from August 4, 2008, to January 30, 2014.
The LAT was operated in sky-survey mode for almost the entire period.
In this observing mode the LAT scans the whole sky, obtaining complete sky coverage
every 2
orbits~($\sim$~3~hr) and approximately uniform exposure.
The analysis was performed over a square
region of 14$^\circ \times$14$^\circ$ with a pixel size of 0$\ddeg$1.
We set the centroid of the region to the SNR center: (R.A., Dec.)~$=$~(34\ddeg42,62\ddeg75)~(J2000).

We used the standard LAT analysis software, the \emph{ScienceTools} version 09-32-05,
publicly available from the \emph{Fermi} Science Support Center (FSSC)\footnote{Software and
documentation of the \emph{Fermi} \emph{ScienceTools} are distributed by \emph{Fermi} Science
Support Center at http://fermi.gsfc.nasa.gov/ssc}.
We used the post-launch instrument response functions~(IRFs) \texttt{P7REP\_SOURCE\_V15}~\citep{IRFs}
 and applied the following event selection
criteria: 
1) events should be classified as reprocessed Pass 7 \emph{Source} class \citep{Atwood09}, 
2) events have a reconstructed zenith angle less than 100$^\circ$, 
to minimize the contamination from Earth-limb gamma-ray emission,
 and 
3) only time intervals when the center of the LAT field of view is within 52$^\circ$ of the local zenith are accepted to further reduce the contamination by Earth's atmospheric emission.
We also eliminated a short period of time (06:25:45 to 06:42:25 UTC on 2012 September 11) during which the LAT detected the bright GeV-emitting GRB~120911B~\citep{GRB} within 15$^\circ$ of HB~3.
{\bf The resulting total exposure was $\sim$~2~$\times$~10$^{11}$~cm$^2$~s~(at 1~GeV).}
We restricted the analysis to the energy range $>200$~MeV to avoid
possible large systematics due to the rapidly varying effective area and much broader PSF at lower energies.

\section{ANALYSIS AND RESULTS}
\label{sec:ana}
\subsection{Morphological analysis}
\label{subsec:spatial}
In order to study the morphology of gamma-ray emission associated with the HB~3 SNR 
we performed a binned likelihood analysis based on Poisson statistics\footnote{As implemented in the publicly available
\emph{Fermi} \emph{Science Tools}.
The documentation concerning the analysis tools and the likelihood fitting procedure is available from
http://fermi.gsfc.nasa.gov/ssc/data/analysis/documentation/Cicerone/.}~\citep[see e.g.][]{Mattox96}.
The likelihood is the product of the probabilities of observing the gamma-ray counts within each spatial and energy bin for a specified emission model.
The probability density function for the likelihood analysis includes (1)~individual sources detected in the 2FGL catalog within 20$^\circ$ of the SNR,
(2)~the Galactic diffuse emission resulting from CR interactions with the interstellar medium and radiation based on the LAT standard diffuse background model, \texttt{gll\_iem\_v05\_rev1.fit} available from the FSSC, 
and (3)~an isotropic component to represent gamma rays and residual CR background using a tabulated spectrum written in \texttt{iso\_source\_v05\_rev1.txt} also available from the FSSC.
We set the spectral parameters of the 2FGL catalog sources located within the SNR~(2FGL J0214.5$+$6251c and 2FGL J0221.4$+$6257c) and within W3~(2FGL J0224.0$+$6204 , 2FGL J0218.7$+$6208c and 2FGL J0225.9$+$6154c) free.  Additionally, the normalizations of the Galactic diffuse and isotropic component were left free.  All other background sources had their indices fixed to the 2FGL values and normalizations set free.
We used only events above 1~GeV~(compared to the 0.2~GeV used in the spectral analysis) for the morphological study to take advantage of the narrower PSF at higher energies.

\begin{figure}
 \plotone{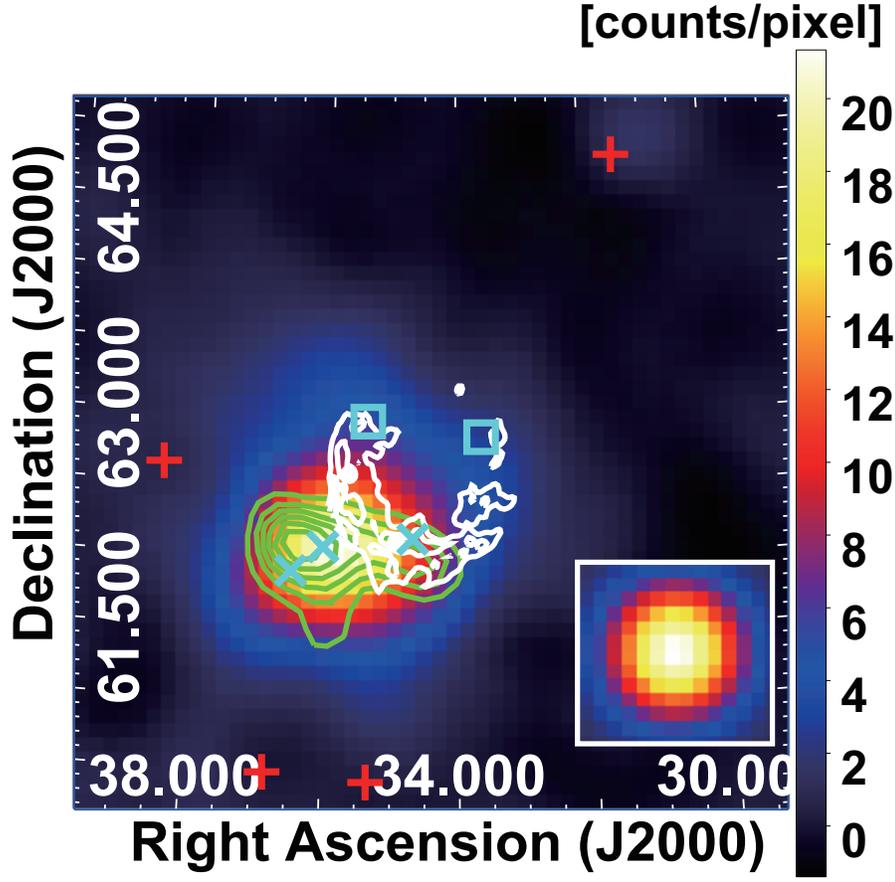}
\caption{
\small
Background-subtracted LAT counts map in the 1--10~GeV energy range around the HB~3 SNR.
The five 2FGL point sources associated with HB~3~(cyan squares) and W3~(cyan crosses) are not included in the background model. 
{\bf The red pluses are the other 2FGL point sources.}
The counts map is binned using a grid of $0\ddeg1$ and smoothed with a Gaussian kernel of $\sigma$ $=$0$\ddeg$25.
The inset of the figure shows the simulated LAT PSF with a photon index
of 2.5 in the same energy range, adopting the same smoothing.
Note that throughout the paper the analysis is conducted on unsmoothed data taking into account the instrument PSF in the likelihood analysis.
{\bf The white contours are 408~MHz radio data~\citep{RadioTemplate} excluding the region around W3, where the emission is predominantly thermal. The contour interval is 2~K from 2~K.
}
Contours of integrated intensity of the  $^{12}$CO~(J=1-0)  line~\citep{CO} are in green.
The contour interval is 2.7~K$\cdot$km~s$^{-1}$ from 2.7~K$\cdot$km~s$^{-1}$.
The CO image was smoothed using a Gaussian kernel with $\sigma$ $=$~0$\ddeg$125.
\label{fig:subtract_cmap}
}
\end{figure}

Figure~\ref{fig:subtract_cmap} shows the counts map in a 4$^\circ~\times$~4$^\circ$ region centered on HB~3, 
after subtracting the Galactic emission, the isotropic component and all the sources from the 2FGL catalog but the five ones spatially coincident with HB~3 and W3.
The image in the {\bf $^{12}$CO~(J=1-0) 
line~\citep{CO}} is overlaid, where the CO line intensities were integrated for velocities with respect to the local standard of rest $-44.2$~km~s$^{-1}$~$<V<-33.8$~km~s$^{-1}$, corresponding to the clouds associated with W3~\citep{COint}.
The correlation between gamma rays and the CO line emission around W3 is evident.

To quantitatively evaluate the correlation with the CO line emission, 
we fitted the LAT counts with the spatial template of the CO line emission instead of the three 2FGL sources within W3.
The shape of CO emission was extracted only around W3. 
Since the edge of the CO emission was unclear due to due to the noise of the spectral measurements,
we varied the threshold value of the CO intensity to extract the shape to maximize the likelihood value with a penalty for one additional parameter.
The spectral shape of W3 was assumed to be a log-parabola function, which is used for the sources corresponding to W3~\citep{2yrCatalog}.
The resulting maximum likelihood values with respect to the maximum likelihood for the null hypothesis~(no emission associated with the SNR and W3) are summarized in Table~\ref{tab:likeratio1}.
A test statistic~(TS) value is defined as $2\ln (L/L_{\rm 0})$, where $L$ and $L_{\rm 0}$ are the maximum likelihoods for the model with/without the source component, respectively~\citep[e.g.][]{Mattox96}.
Although the spatial models are not nested so there is no rigorous way to directly compare them,
the TS value for the CO image is significantly larger than for the three 2FGL sources and is physically motivated.

\begin{figure}
 \plotone{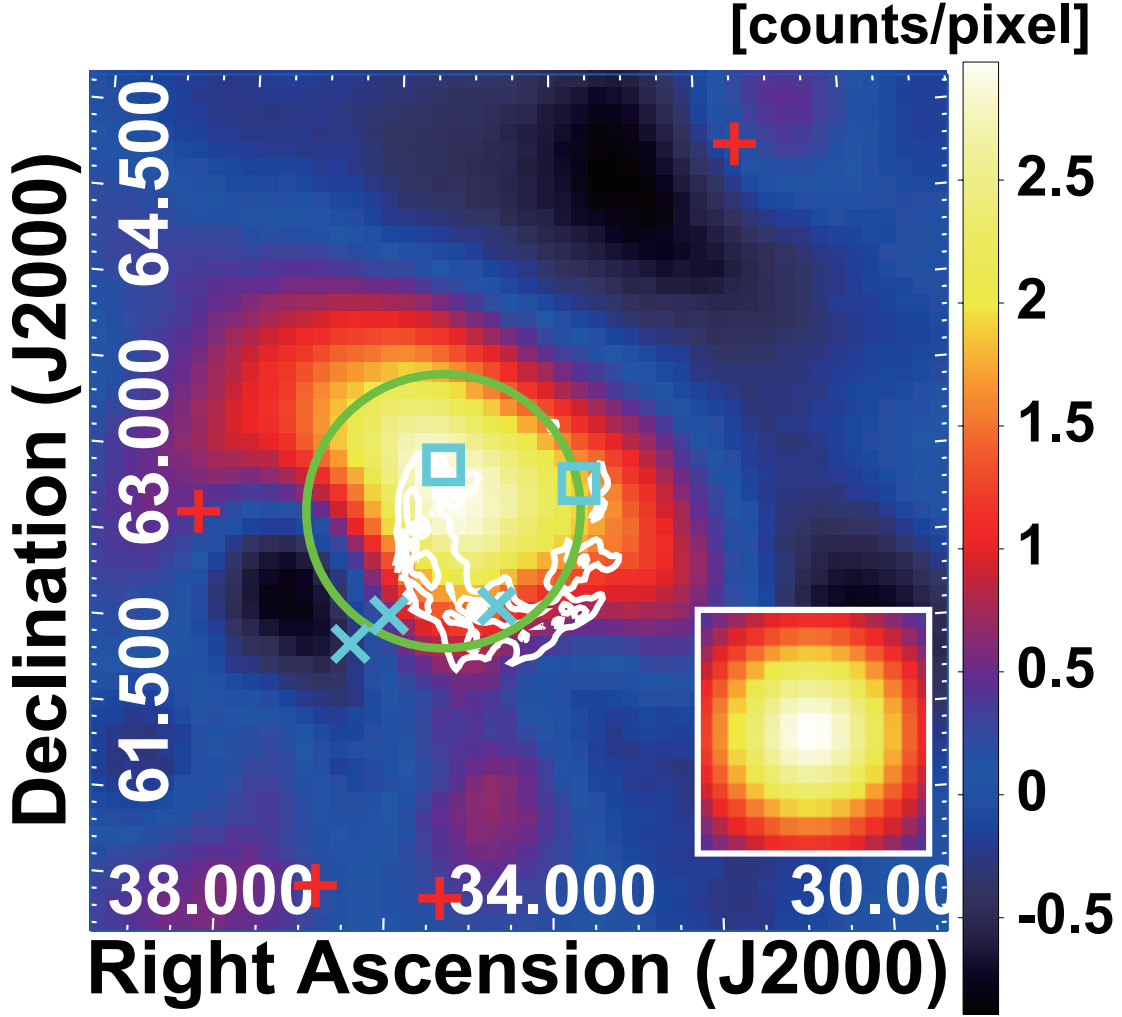}
\caption{
Background-subtracted LAT counts map in the 1--10~GeV energy range.
The two LAT point sources associated with HB~3 are not included in the background model, but the CO template for W3 is. 
The counts map is binned using a grid of $0\ddeg1$ and smoothed with a Gaussian kernel of sigma $=$~0$\ddeg$40.
The green circle indicates the best-fit disk for HB~3 used for the analysis as a spatial template.
The details of the other overlays are described in the caption of Figure~\ref{fig:subtract_cmap}.
\label{fig:subtract_cmap2}
}
\end{figure}

We further characterized the morphology of gamma-ray emission associated 
with the SNR. Figure~\ref{fig:subtract_cmap2} shows the counts map in a 4$^\circ~\times$~4$^\circ$ region centered on HB~3,
after subtracting the Galactic emission, the isotropic component, the 2FGL point sources except for the 2 sources associated with the SNR, and W3 modeled by the CO template.
This highlights the diffuse emission around the SNR.
To quantitatively evaluate the detection significance of the diffuse emission from HB~3,
we adopted a uniform disk as a spatial template assuming a simple power-law spectrum.
We varied the radius and location of the disk and evaluated the maximum likelihood values.
The best-fit disk has the radius of $0\ddeg 80^{+0.18}_{-0.17}$ centered on (R.A., Dec.)~$=$~(35\ddeg36, 62\ddeg69)~(J2000).
The error of the centroid is 0$\ddeg$06 at 68\% confidence level.
We note that the maximum likelihood value for a 408~MHz radio template with suppression of emission from W3 was significantly worse than the best-fit disk model, indicating that the gamma-ray emission around HB~3 is not strongly spatially associated with the shock region of HB~3 as traced by radio.
{\bf The best-fit disk is offset from the SNR center (R.A., Dec.) $=$ (34\ddeg51, 62\ddeg79)~(J2000) \citep{SNRcenter} by 0\ddeg40, and the offset is significant at the 4.2$\sigma$ level.}
The TS value for the best-fit disk is larger than for the two 2FGL sources. 
In addition, the detection significance for the disk with energies of $>$~1~GeV is at the $\sim$~12~$\sigma$ level.
Therefore we finally adopted the uniform disk template with maximum likelihood parameters for the whole SNR in the following spectral analysis.

\subsection{Spectral analysis} 
\begin{figure}
\plotone{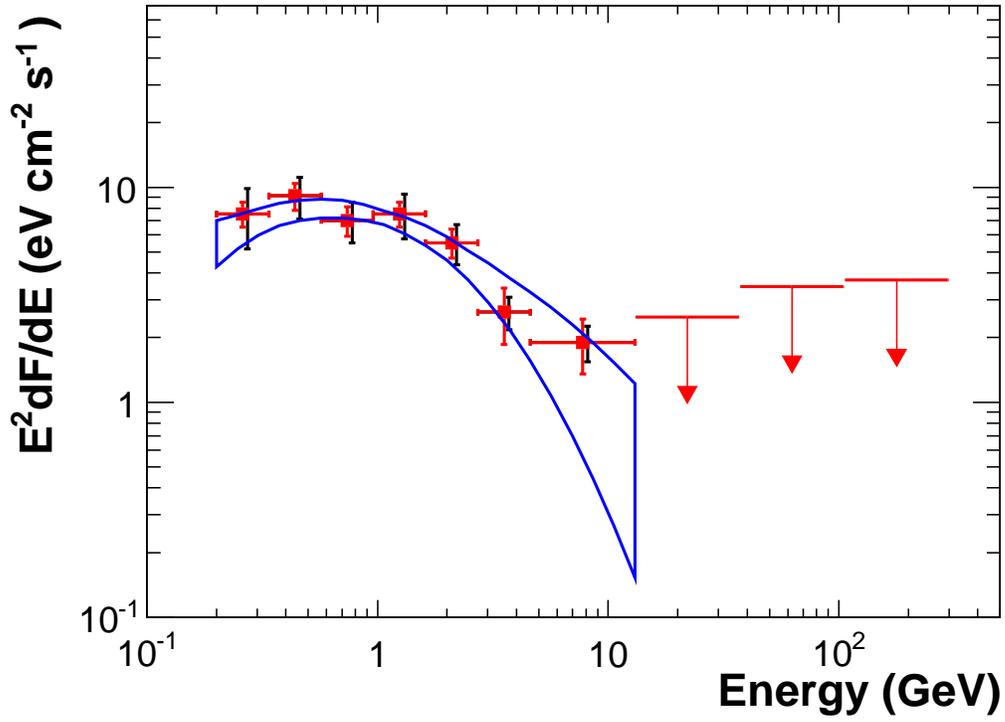}
\caption{ Spectral energy distribution of the gamma-ray emission measured by the LAT for the HB~3 SNR.
Red squares are LAT flux points.
Horizontal bars indicate the energy range the flux refers to.
Vertical bars show statistical errors in red and systematic errors
 in black.
In energy bins where
the detection is not significant (TS $<$~4) we show upper limits at the 2 sigma confidence level.
The blue region is the 68~\% confidence range~(no systematic error) of the LAT spectrum assuming that the spectral shape is a log parabola.
 \label{fig:spec}}
\end{figure}

\begin{figure}
 \plotone{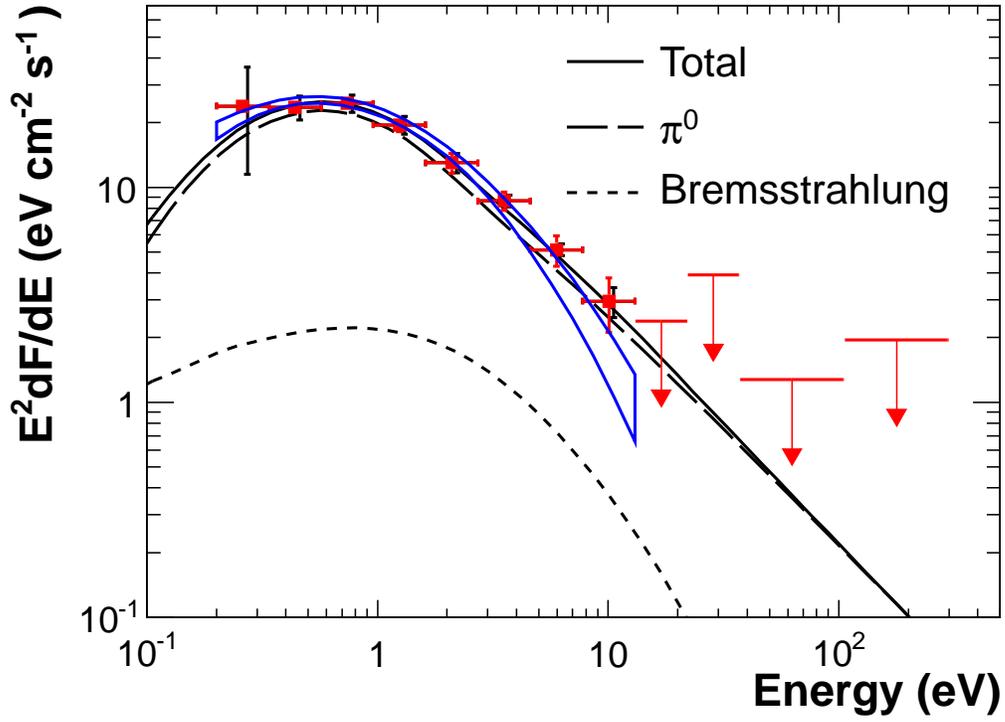}
\caption{ Spectral energy distribution of the gamma-ray emission measured by the LAT for W3. 
The details of the plots and bars are described in the caption of Figure~\ref{fig:spec}.
The blue region is the 68~\% confidence range~(no systematic error) of the LAT spectrum assuming that the spectral shape is a log parabola.
The lines shows a $\pi^0$-decay dominated model.
The details of the model lines are described in the caption of Figure~\ref{fig:spec_multi}.
 \label{fig:spec_w3}}
\end{figure}

To measure the spectra for the HB~3 SNR and W3, we performed maximum likelihood fits in logarithmically-spaced energy bands from 0.2~GeV to 300~GeV, 
using the spatial model described above with the spectral indices in each bin fit to 2.
Figure~\ref{fig:spec} and \ref{fig:spec_w3} show the resulting spectral energy distributions~(SEDs).
Upper limits at the 2 sigma confidence level are calculated assuming a photon
index of 2 if the detection is not significant in an energy bin, i.e., the TS value with respect to the null hypothesis is less than 4~(corresponding to 2~$\sigma$
for one additional degree of freedom).

We identify at least four different sources of systematic uncertainties affecting the
estimate of the fluxes. Uncertainties in the LAT effective area are evaluated by comparing the efficiencies of analysis cuts for data and simulation of observations of 
Vela and the limb of the Earth, among other consistency checks~\citep{IRFs}. For \texttt{P7REP\_SOURCE\_V15}, these studies suggest a 10\% 
systematic uncertainty below 100~MeV, decreasing linearly with the logarithm of energy to 5\% in the range between 316~MeV and 10~GeV and 
increasing linearly with the logarithm of energy up to 15\% at 1~TeV~\citep{IRFs}.
We estimated the systematic errors by changing the effective area of the IRFs energy-dependently considering the above uncertainties.

We adopted the strategy described in \cite{diffusesys1} and \cite{diffusesys2} to evaluate the systematic uncertainties due to the modeling of interstellar emission.
We compared the results obtained using the standard model in Section~\ref{subsec:spatial} with the results based on eight alternative interstellar emission models. We varied the uniform spin temperature used to estimate the column densities of interstellar atomic hydrogen, the vertical height of the CR propagation halo, and the CR source distribution in the Galaxy.
We similarly gauged the uncertainties due to the morphological template by comparing the results with those obtained by changing the radius of the disk template by $\pm$~1$\sigma$.
Lastly, we obtained the uncertainties due to the assumption of the photon indices. 
We changed the indices from 0 to 4.
The total systematic errors shown in Figure~\ref{fig:spec} and \ref{fig:spec_w3} are set by adding the above uncertainties in quadrature. 

We probed for a spectral break in the LAT energy band by comparing the likelihood values of a spectral fit over the whole energy range considered based on a simple power law and a log parabola function.
The TS values and best-fit parameters are summarized in Table~\ref{tab:spectral_shape}.
The TS values for the log-parabola function correspond to improvements at the $> 4$~$\sigma$ confidence level for the SNR and $> 16$~$\sigma$ for W3, respectively.
Assuming the spectral shape can be described by a log parabola function,
the gamma-ray energy fluxes between 0.2--300~GeV inferred from our analysis are $3.7 \times 10^{-11}$erg~cm$^{-2}$~s$^{-1}$ for the SNR and $1.0 \times 10^{-10}$erg~cm$^{-2}$~s$^{-1}$ for W3, respectively.
The gamma-ray luminosities between 0.2--300~GeV assuming the distance is 2.2~kpc are $2.1 \times 10^{34}$erg~s$^{-1}$ for the SNR and $5.9 \times 10^{34}$erg~s$^{-1}$ for W3, respectively.

\section{DISCUSSION}
\label{sec:discuss}
There is a correspondence between gamma-ray emission and the SNR.
{\bf
The gamma-ray emission around HB 3 is extended, located off the Galactic plane and far from the Galactic center, so the probability of chance superposition of other Galactic sources is small.
Also, the spectral shape of gamma rays around HB~3 is similar to other gamma-ray sources associated with old SNRs, i.e. a soft spectrum with a spectral break at energies of around 1--10~GeV~\citep[e.g.,][]{W51C,W44,IC443,W28}.
These facts suggest the association between the gamma-ray emission and HB~3.
}
Assuming this, enhancement of gamma-ray emission from W3, which contains the molecular clouds traced by CO line emission, can be reasonably explained by interaction between high-energy particles and the dense material in the clouds.
Thus we argue that the bulk of gamma-ray emission comes from interactions of high-energy particles accelerated at the shocks of HB~3 with interstellar matter or fields in the regions.
First, we discuss the gamma-ray spectrum of the SNR which was obtained by using the disk spatial template.
To model the broadband emission from the entire SNR we adopt the simplest possible assumption that gamma rays are emitted by a population of accelerated protons and electrons distributed in the same region 
and characterized by constant matter density and magnetic field strength. 
We assume the injected electrons to have the same momentum distribution as protons.
This assumption requires a break in the momentum spectrum
because the spectral index in the radio domain, corresponding to lower
particle momenta, is much harder than for gamma rays, which
correspond to higher particle momenta.
Therefore, we use the following functional form to model the momentum distribution of injected
particles:
\begin{equation}
Q_{e, p}(p) = a_{e,p} \left( \frac{p}{1~{\rm GeV}~c^{-1}} \right)^{-s_{\rm L}} \left\{ 1+\left(\frac{p}{p_{\rm br}} \right)^2 \right\} ^{-(s_{\rm H}-s_{\rm L})/2},
\end{equation}
 where $p_{\rm br}$ is the break momentum, $s_{\rm L}$ is the spectral index below the break
and $s_{\rm H}$ above the break.
Note that here we consider a minimum momenta
of 100~MeV~$c^{-1}$ since the details of the proton/electron injection process are poorly known.

Electrons suffer energy losses due to ionization, Coulomb scattering, bremsstrahlung,
synchrotron
emission and inverse Compton~(IC) scattering.
We calculated the evolution of the electron momenta spectrum by the following equation:
\begin{equation}
  \frac{\partial N_{e,p}}{\partial t} = \frac{\partial}{\partial p} \left( b_{e,p}N_{e,p} \right) + Q_{e,p} ,
\end{equation}
where $b_{e,p} = -dp/dt$ is the momentum loss rate, and $Q_{e,p}$ is the particle injection rate.
We assume $Q_{e,p}$ to be constant, i.e., that the  
shock produces a constant number of particles 
until the SNR enters the radiative phase, at which time the source turns off. 
This prescription approximates the weakening of the shock and the reduction in the particle acceleration efficiency, which would be properly treated by using a time-dependent shock compression ratio~\citep{SNRaccelerationTime}. 
To derive the remnant emission spectrum we calculated $N_{e,p}(p,T_{\rm 0})$
numerically, where $T_{\rm 0}$ is the SNR age assumed to be $3 \times 10^4$~yr.
Note that we neglected the momentum losses for protons since the timescale of neutral pion production is $\sim$~10$^7/\nh$~yr where $\nh$ is the gas density averaged over the entire SNR shell and is much longer than the SNR age.
Also we neglect the gamma-ray emission by secondary positrons and electrons from
charged pion decay, 
because the emission from secondaries is generally faint compared to that from primary electrons unless the gas density is as high as that in dense molecular clouds 
and the SNR evolution reaches the later stages,
or the injected electron-to-proton ratio is much lower than locally observed.
The gamma-ray spectrum from $\pi^0$ decay produced by the interactions of protons with
ambient hydrogen is calculated based on \cite{Dermer86}
using a scaling factor of 1.68 to account for helium and heavier nuclei in target
material and cosmic rays~\citep{Mori09}. 
Contributions from bremsstrahlung and inverse Compton scattering by accelerated electrons
are computed based on \cite{Blumenthal70}, whereas synchrotron radiation is based on \cite{Crusius86}.

\begin{figure}
 \plotone{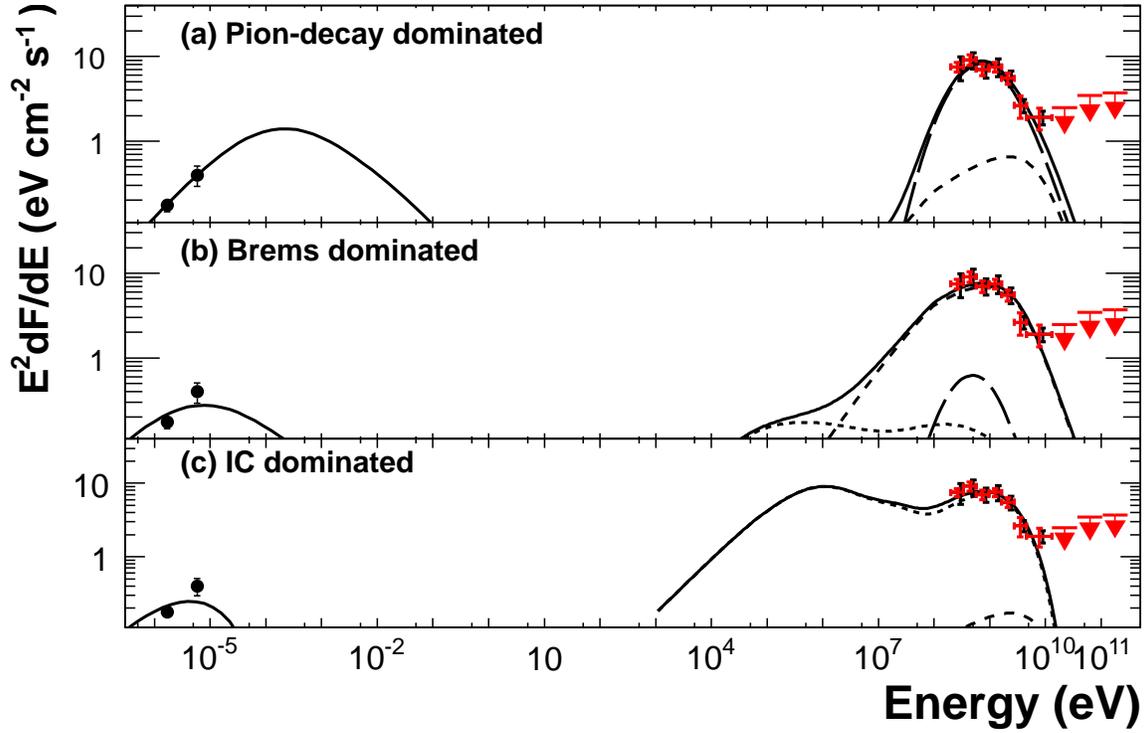}
\caption{Multi-band spectrum of the HB~3 SNR.
 \label{fig:spec_multi}
In the GeV band LAT measurements are reported as in Figure~\ref{fig:spec}.
The radio continuum emission~\citep{RadioSpectrum} is shown by black dots. Radio emission is modeled as synchrotron radiation, 
while gamma-ray emission is modeled by different combinations of $\pi^0$-decay~(long-dashed curve),
bremsstrahlung~(dashed curve), and inverse Compton~(IC) scattering~(dotted curve). 
Details of the models are
described in the text and numerical values are given in Table~\ref{tab:model}: a) $\pi^0$-decay dominated model, b)
bremsstrahlung dominated model, c) IC dominated model.}
\end{figure}

First, we consider a $\pi^0$-decay dominated model.
The number index of protons in the high-energy regime
 is constrained to be $s_{\rm H} \approx 4.0$   from the gamma-ray spectral slope.
$s_{\rm L} \approx 1.5$ by modeling the radio spectrum as synchrotron radiation by relativistic
electrons (under the assumption that protons and electrons have identical injection spectra). The spectral index $\alpha$ of the radio continuum emission is $\sim$~0.34~\citep{RadioSpectrum}, where $\alpha$ is defined as $S_\nu \propto \nu^{-\alpha}$ where $S_\nu$ and $\nu$ are the flux density and the frequency, respectively.
The gamma-ray spectrum provides an upper bound for the momentum break at $\sim$~ 13~GeV~$c^{-1}$.
We adopt a break at the best-fit value, 10~GeV~$c^{-1}$.
The gas density is fixed to $2.0~\mathrm{cm}^{-3}$ based on the \hi observations~\citep{density}.
The resulting total proton energy,  $W_{p}\sim 5.2~\times~10^{49}~\cdot~(2.0~\mathrm{cm}^{-3}/\nh)\cdot(d/2.2 \mathrm{kpc})^{2}$~erg, is less than 10~\% 
of the typical kinetic energy of a supernova explosion.
For an electron-to-proton ratio $K_{ep} = 0.01$ at 1~GeV~$c^{-1}$, which
is the ratio measured at the Earth,
the magnetic field strength is constrained to be $B \sim 25\;\upmu$G by radio data.
Using the parameters summarized in Table~\ref{tab:model}, 
we obtained the SEDs shown in Figure~\ref{fig:spec_multi}~(a).

The gamma-ray spectrum can be reproduced by an electron bremsstrahlung dominated model 
in Figure~\ref{fig:spec_multi}~(b), although it is not easy to fit the data in the radio band simultaneously.
The density used in this model is the same as that for the $\pi^0$-decay dominated model, while $K_{ep}$ is set to 1 so that the leptonic emission dominates.

An inverse Compton dominated model can also reproduce the gamma-ray spectrum as shown in Figure~\ref{fig:spec_multi}~(c).
Gamma-ray emission of IC origin is
 due to interactions of high-energy electrons with optical and infrared radiation fields and the cosmic microwave background~(CMB).
We used in our calculations the first two components as they are modeled in~\citet{Porter08} at the location of HB~3.
Since their spectra are very complex, they are approximated by two
infrared blackbody components~($T_{\rm IR} = 33.3, 1.00 \times 10^3$~K, $U_{\rm IR} = 0.17, 0.04$~eV~cm$^{-3}$, respectively),
and 
two optical blackbody components~($T_{\rm opt} = 4.02 \times 10^3, 1.34 \times 10^4$~K, $U_{\rm opt} = 2.71 \times 10^{-1}, 8.80 \times 10^{-2}$~eV~cm$^{-3}$, respectively). 
The flux ratio between the IC and the synchrotron components 
constrains the magnetic field to be less than 1~$\upmu$G
and requires a low gas density of 
 $\nh \sim 2.0 \times 10^{-3}~\mathrm{cm}^{-3}$ to suppress the electron bremsstrahlung,
which is unlikely.
Increasing the intensity of the interstellar radiation field other than CMB
would loosen the constraint on the magnetic field and the gas density.
A radiation field about 100 times more intense is required to satisfy the above assumption on the magnetic field and the gas density.


Assuming that $\pi^0$-decay produced by the interactions of protons with ambient hydrogen is responsible for the bulk of the gamma-ray emission from the SNR,
the gamma-ray spectrum of W3 can be reasonably explained by a $\pi^0$-decay dominated model. If the protons are accelerated in the whole SNR in the same manner and are not largely affected by spectral deformation due to cosmic-ray diffusion process, the gamma-ray spectrum of W3 is expected to be modeled with a similar proton spectrum. 
The curved lines in Figure~\ref{fig:spec_w3} show the gamma-ray spectrum of W3 with the $\pi^0$-decay dominated model assuming the density in the molecular clouds is $100~\mathrm{cm}^{-3}$. 
The spectrum can be reproduced without any change from the proton momentum spectrum of the SNR.
The resulting total proton energy,  $W_{p}\sim 0.29~\times~10^{49}~\cdot~(100~\mathrm{cm}^{-3}/\nh)\cdot(d/2.2 \mathrm{kpc})^{2}$~erg, is about 5.5~\% 
of that for the $\pi^0$-decay model of the SNR, which is likely considering the solid angle of W3 from the center of the SNR.
{\bf We also estimated the energy density of cosmic rays in W3 in order to confirm the enhancement of the energy density due to the SNR
compared to that from the local cosmic rays. The mass of the whole clouds of W3 is $4.4 \times 10^5~\masssun$~\citep{W3COmass}. 
We assume the cloud mass related to the gamma-ray emission to be about $10^4~\masssun$ since the gamma-ray emission only comes from a small part of the W3 giant molecular cloud.
The resulting density is $15\cdot(10^4 \masssun/ \mass)$~eV~cm$^{-3}$, where $\mass$ is the cloud mass related to the gamma-ray emission. This is much higher than the energy density of the local cosmic rays, indicating a strong association between the gamma-ray emission and the SNR.
}
To summarize, it is most natural to assume that gamma-ray emission from HB~3 is dominated by decay of $\pi^0$ produced in nucleon-nucleon interactions of hadronic cosmic rays with interstellar matter.
It should be emphasized that our observations around HB~3 provide a rare and valuable example where there are detections of gamma rays from both the adjacent interacting molecular cloud and the SNR itself.
{\bf 
This differs from W44~\citep{W44second} and W28~\citep{W28second} where the adjacent gamma-ray emitting clouds are thought to be outside the SNRs.}

\section{CONCLUSIONS}
\label{sec:conclusion}
We detected extended gamma-ray emissions by using the LAT data in the region of the SNR HB~3 and the W3 \hii complex.
The decay of $\pi^0$ produced by interactions of hadrons accelerated
by the remnant with interstellar gas naturally explains the gamma-ray emission around HB~3. 
Assuming this, the gamma-ray emission from W3 is also reasonably explained by the interaction of cosmic rays accelerated by the SNR with the dense gas in the clouds associated with W3.

\acknowledgments
The \textit{Fermi} LAT Collaboration acknowledges generous ongoing support
from a number of agencies and institutes that have supported both the
development and the operation of the LAT as well as scientific data analysis.
These include the National Aeronautics and Space Administration and the
Department of Energy in the United States, the Commissariat \`a l'Energie Atomique
and the Centre National de la Recherche Scientifique / Institut National de Physique
Nucl\'eaire et de Physique des Particules in France, the Agenzia Spaziale Italiana
and the Istituto Nazionale di Fisica Nucleare in Italy, the Ministry of Education,
Culture, Sports, Science and Technology (MEXT), High Energy Accelerator Research
Organization (KEK) and Japan Aerospace Exploration Agency (JAXA) in Japan, and
the K.~A.~Wallenberg Foundation, the Swedish Research Council and the
Swedish National Space Board in Sweden.
 
Additional support for science analysis during the operations phase is gratefully
acknowledged from the Istituto Nazionale di Astrofisica in Italy and the Centre National d'\'Etudes Spatiales in France.

The research presented in this paper has used data from the Canadian Galactic Plane Survey, a Canadian project with international partners, supported by the Natural Sciences and Engineering Research Council.

{\bf We wish to thank Dave Green~(University of Cambridge) for advising us about the reference for the position of the SNR.
}

\begin{table}
\begin{center}
\caption{ Test Statistics for Different Spatial Models Compared with the 
Null hypothesis~(no gamma-ray emission associated neither with HB~3 nor W3) in the 1--300~GeV energy range.
\label{tab:likeratio1}}
\begin{tabular}{lccc}
\tableline\tableline
 Model   & Test Statistic &  Additional Degrees of Freedom  \\
\tableline
   3 point sources $+$ 2 point sources\tablenotemark{a} & 2025.5 &  13 \\
   CO image  \tablenotemark{b} $+$ 2 point sources\tablenotemark{c} & 2115.8 &  9 \\
   CO image $+$ Uniform disk\tablenotemark{d} & 2142.2 &  9 \\
   CO image $+$ Radio template\tablenotemark{e} & 2111.8 & 6 \\
\tableline
\end{tabular}
\tablenotetext{\rm a}{The five sources listed in the 2FGL source list associated with the HB~3 SNR and the W3~\citep{2yrCatalog}.
The spectral shapes are a power-law function for 2 sources, a log parabola function for 3 sources.  }
\tablenotetext{\rm b}{The additional degrees of freedom for the CO image are 2 for the spectral shape, 1 for the threshold value of the CO intensity to extract the shape.
The threshold value to maximize the likelihood value is 3~K$\cdot$km~s$^{-1}$.
The details are shown in the text. }
\tablenotetext{\rm c}{The two sources listed in the 2FGL source list associated with HB~3.}
\tablenotetext{\rm d}{The additional degrees of freedom for the uniform disk are 5, 2 for the spectral shape~(a power law), 3 for the disk radius and position.
}
\tablenotetext{\rm e}{The radio template was obtained from 408~MHz radio data~\citep{RadioTemplate} by excluding the region around W3, where the emission is predominantly thermal.
The additional degrees of freedom for the uniform disk are 2 for the spectral shape (a power law).}
\end{center}
\end{table}

\begin{table}
\begin{center}
\caption{Test Statistics and Parameters for Spectral Models~(0.2--300~
GeV) \label{tab:spectral_shape}}
\begin{tabular}{lcccc}
\tableline\tableline
 Spectral Model  & Test Statistic\tablenotemark{a}  &  Degrees  & Spectral Parameters  \\  &  &   of Freedom & \\
\tableline\tableline
HB~3 &   &  &  \\
\tableline
Power Law  & 0  & 2 & $E^{-p}$; $p=2.39\pm0.05$ \\
\tableline
Log Parabola  & 18 &  3 &   $\left(\frac{E}{1~{\rm GeV}}\right)^{-p_1-p_2\log{\left(\frac{E}{1~{\rm GeV}}\right)}}$ \\
 & & &  $p_1=2.29 \pm 0.09$ \\
 & & &  $p_2=0.30 \pm 0.11$ \\
\tableline\tableline
W3 &   &  &  \\
\tableline
Power Law  & 0  & 2 & $E^{-p}$; $p=2.42\pm0.02$ \\
\tableline
Log Parabola  & 267 &  3 &   $\left(\frac{E}{1~{\rm GeV}}\right)^{-p_1-p_
2\log{\left(\frac{E}{1~{\rm GeV}}\right)}}$ \\
 & & &  $p_1=2.40 \pm 0.03$ \\
 & & &  $p_2=0.33 \pm 0.03$ \\
\tableline\tableline
\end{tabular}
\tablenotetext{\rm a}{$2\ln (L/L_{\rm 0})$, where $L$ and $L_{\rm 0}$ are the maximum likelihood values for the 
model under consideration and the power-law model, respectively.}
\end{center}
\end{table}

\begin{table}
\begin{center}
\caption{Model parameters for the HB~3 SNR and W3.\label{tab:model}}
\begin{tabular}{lccccccccc}
\tableline\tableline
 HB~3 Model  & \kep\tablenotemark{a} &  $s_{\rm L}$\tablenotemark{b} & $p_{\rm br}$\tablenotemark{c} & $s_{\rm H}$\tablenotemark{d} & $B$ &
 $\nh$\tablenotemark{e} & $W_{p}$\tablenotemark{f} & $W_{e}$\tablenotemark{f} \\
   &    &  & (GeV~$c^{-1}$) & &  ($\upmu$G) & (cm$^{-3}$) & ($10^{49}$~erg) & ($10^{49}$~erg) \\
\tableline
(a)~Pion & 0.01  & 1.5 & 10 & 4.0 & 25 & 2.0 & 5.2 & 0.074 \\
(b)~Bremsstrahlung & 1 & 1.5 & 4.0 & 4.0 & 5.0 & 2.0  & 0.46 & 0.84  \\
(c)~Inverse Compton & 1 & 1.5 & 60 & 40 & 0.64 & $2.0 \times 10^{-3}$ & 12 & 18 \\
(d)~Inverse Compton & 1 & 1.5 & 60 & 40 & 15 & 2.0 & 0.11 & 0.16 \\
~~(photon field $\times$ 100) & & & & & & & & \\
\tableline\tableline
 W3 Model & & & & & & & & \\
\tableline
Pion & 0.01  & 1.5 & 10 & 4.0 & $\verb|<|$ 200\tablenotemark{g} & 100 & 0.29 & $3.4 \times 10^{-3}$ \\
\tableline
\end{tabular}
\tablenotetext{\rm a}{The ratio electrons-to-protons at 1~GeV~$c^{-1}$.}
\tablenotetext{\rm b}{The momentum distribution of particles is assumed to be a smoothly broken power law, where the indices and the break
 momentum are identical for both accelerated protons and electrons. $s_{\rm L}$ is the spectral index in momentum below the break.}
\tablenotetext{\rm c}{$p_{\rm br}$ is the break momentum.}
\tablenotetext{\rm d}{Spectral index in momentum above the break.}
\tablenotetext{\rm e}{Average hydrogen number density of ambient medium.}
\tablenotetext{\rm f}{
The total energy is calculated for particles $>$~100~MeV~$c^{-1}$.}
\tablenotetext{\rm g}{This upper limit was obtained by assuming the flux of the non-thermal radio emission around W3 is the same as that from the whole SNR since the radio emission from W3 is predominantly thermal.}
\end{center}
\end{table}


\begin{thebibliography}{}


\bibitem[Abdo et al.(2009a)]{onorbit_cal} Abdo, A.~A., Ackermann, M., Ajello, M., et al.\ 2009, Astroparticle Physics, 32, 193 

\bibitem[Abdo et al.(2009b)]{W51C} Abdo, A.~A., Ackermann, M., Ajello, M., et al.\ 2009, \apjl, 706, L1 

\bibitem[Abdo et al.(2010a)]{W44} Abdo, A.~A., Ackermann, M., Ajello, M., et al.\ 2010, Science, 327, 1103 

\bibitem[Abdo et al.(2010b)]{IC443} Abdo, A.~A., Ackermann, M., Ajello, M., et al.\ 2010, \apj, 712, 45

\bibitem[Abdo et al.(2010c)]{W28} Abdo, A.~A., Ackermann, M., Ajello, M., et al.\ 2010, \apj, 718, 348


\bibitem[Ackermann et al.(2012)]{IRFs} Ackermann, M., Ajello, M., Albert, A., et al.\ 2012, \apjs, 203, 4 


\bibitem[Ackermann et al.(2013)]{diffusesys1} Ackermann, M., 
Ajello, M., Allafort, A., et al.\ 2013, Science, 339, 807 

\bibitem[Atwood et al.(2009)]{Atwood09} Atwood, W. B., et al. (The \emph{Fermi} LAT Collaboration) 2009, \apj, 697, 1071 


\bibitem[Blandford \& Eichler(1987)]{blandford87} Blandford, R. D. \& Eichler, D. 1987, Phys. Rep., 154, 1

\bibitem[Blumenthal \& Gould(1970)]{Blumenthal70} Blumenthal, G.~R., \& Gould, R.~J.\ 1970, Reviews of Modern Physics, 42, 237 

\bibitem[Crusius \& Schlickeiser(1986)]{Crusius86} Crusius, A., \& Schlickeiser, R.\ 1986, \aap, 164, L16

\bibitem[Dame et al.(2001)]{CO} Dame, T.~M., Hartmann, D., \& Thaddeus, P.\ 2001, \apj, 547, 792 

\bibitem[de Palma et al.(2013)]{diffusesys2}
de Palma, F., Brandt, T. J., Johannesson, G., \& Tibaldo, L. Fermi LAT Collaboration 2013, Proc. of 4th International Fermi Symposium Proc., ed. N. Omodei, G. Senatore, T. Brandt, \& C. Wilson-Hodge (Stanford, CA: Stanford Univ.), 172

\bibitem[Dermer(1986)]{Dermer86} Dermer, C.~D.\ 1986, \aap, 157, 223 

\bibitem[Gosachinskii(2005)]{density} Gosachinskii, I.~V.\ 2005, Astronomy Letters, 31, 179

\bibitem[Hanabata et al.(2014)]{W28second} Hanabata, Y., Katagiri, H., Hewitt, J.~W., et al.\ 2014, \apj, 786, 145 

\bibitem[Hartman et al.(1999)]{3eg} Hartman, R.~C., 
Bertsch, D.~L., Bloom, S.~D., et al.\ 1999, \apjs, 123, 79 


\bibitem[Green(2014)]{newRadioShape} Green, D.~A.\ 2014, Bulletin of the Astronomical Society of India, 42, 47 


\bibitem[Hartman et al.(1999)]{3eg} Hartman, R.~C., Bertsch, D.~L., Bloom, S.~D., et al.\ 1999, \apjs, 123, 79 

\bibitem[Kothes et al.(2006)]{SNRcenter} Kothes, R., Fedotov, K., Foster, T.~J., \& Uyan{\i}ker, B.\ 2006, \aap, 457, 1081 


\bibitem[Lazendic \& Slane(2006)]{age} Lazendic, J.~S., \& Slane, P.~O.\ 2006, \apj, 647, 350 

\bibitem[Mattox et al.(1996)]{Mattox96} Mattox, J. R., et al. , 1996, \apj, 461, 396

\bibitem[Moraal \& Axford(1983)]{SNRaccelerationTime} Moraal, H., \& Axford, W.~I.\ 1983, \aap, 125, 204 


\bibitem[Mori(2009)]{Mori09} Mori, M., 2009, Astropart. Phys., 31, 341


\bibitem[Nolan et al.(2012)]{2yrCatalog} Nolan, P.~L., Abdo, A.~A., Ackermann, M., et al. (The \emph{Fermi} LAT Collaboration) 2012, \apjs, 199, 31 





\bibitem[Protassov et al.(2002)]{TSconversion} Protassov, R., van Dyk, D.~A., Connors, A., Kashyap, V.~L., \& Siemiginowska, A.\ 2002, \apj, 571, 545


\bibitem[Routledge et 
al.(1991)]{COint} Routledge, D., Dewdney, P.~E., Landecker, T.~L., \& Vaneldik, J.~F.\ 1991, \aap, 247, 529 

\bibitem[Polychroni et al.(2012)]{W3COmass} Polychroni, D., 
Moore, T.~J.~T., \& Allsopp, J.\ 2012, \mnras, 422, 2992 

\bibitem[Porter et al.(2008)]{Porter08} Porter, T., et al. 2008, \apj, 682, 400

%
\bibitem[Racusin et al.(2012)]{GRB} Racusin, J.~L., Vianello, G., Kocevski, D., Omodei, N., \& Ohno, M.\ 2012, GRB Coordinates Network, 13756, 1 


\bibitem[Strong(1977)]{COSB} Strong, A.~W.\ 1977, \nat, 269, 394 


\bibitem[Taylor et al.(2003)]{RadioTemplate} Taylor, A.~R., Gibson, S.~J., Peracaula, M., et al.\ 2003, \aj, 125, 3145 

\bibitem[Tian \& Leahy(2005)]{RadioSpectrum} Tian, W.~W., \& Leahy, D.\ 2005, \aap, 436, 187 

\bibitem[Thompson et al.(2012)]{SNRref} Thompson, D.~J., Baldini, L., \& Uchiyama, Y.\ 2012, Astroparticle Physics, 39, 22 

\bibitem[Uchiyama et al.(2012)]{W44second} Uchiyama, Y., Funk, S., Katagiri, H., et al.\ 2012, \apjl, 749, L35 

\bibitem[Xu et al.(2006)]{dist4} Xu, Y., Reid, M.~J., Zheng, X.~W., \& Menten, K.~M.\ 2006, Science, 311, 54 


\end{thebibliography}
\end{document}